\documentclass [a4paper,UKenglish]{lipics-v2016}
\usepackage[utf8]{inputenc}
\usepackage{cite}
\usepackage{amssymb}
\usepackage{amsmath}
\usepackage{graphicx}
\usepackage[ruled,vlined,lined,boxed,commentsnumbered]{algorithm2e}

\newcommand{\bigO}{\mathcal O}

\usepackage{microtype}

\title{Practical and Effective Re-Pair Compression}

\author[1]{Philip Bille}
\author[1]{Inge Li G{\o}rtz}
\author[1]{Nicola Prezza}
\affil[1]{Technical University of Denmark, DTU Compute, \\\texttt{phbi@dtu.dk, inge@dtu.dk, npre@dtu.dk}}

\authorrunning{P. Bille, I.L. G{\o}rtz, and N. Prezza} 

\Copyright{Philip Bille, Inge L. G{\o}rtz, and Nicola Prezza}


\EventEditors{John Q. Open and Joan R. Acces}
\EventNoEds{2}
\EventLongTitle{42nd Conference on Very Important Topics (CVIT 2016)}
\EventShortTitle{CVIT 2017}
\EventAcronym{CVIT}
\EventYear{2017}
\EventDate{December 24--27, 2016}
\EventLocation{Little Whinging, United Kingdom}
\EventLogo{}
\SeriesVolume{XX}
\ArticleNo{YY}

\begin{document}

\maketitle


\begin{abstract}
	Re-Pair is an efficient grammar compressor that operates by recursively replacing high-frequency character pairs with new grammar symbols. 
	The most space-efficient linear-time algorithm computing Re-Pair uses $(1+\epsilon)n+\sqrt n$ words on top of the re-writable text (of length $n$ and stored in $n$ words), for any constant $\epsilon>0$; in practice however, this solution uses complex sub-procedures preventing it from being practical. 
	In this paper, we present an implementation of the above-mentioned result making use of more practical solutions; our tool further improves the working space to $(1.5+\epsilon)n$ words (text included), for some small constant $\epsilon$. 
	As a second contribution, we focus on compact representations of the output grammar. The lower bound for storing a grammar with $d$ rules is $\log(d!)+2d\approx d\log d+0.557 d$ bits, and the most efficient encoding algorithm in the literature uses at most $d\log d + 2d$ bits and runs in $\bigO(d^{1.5})$ time. We describe a linear-time heuristic maximizing the compressibility of the output Re-Pair grammar. On real datasets, our grammar encoding uses---on average---only $2.8\%$ more bits than the information-theoretic minimum. In half of the tested cases, our compressor improves the output size of \texttt{7-Zip} with maximum compression rate turned on.
\end{abstract}


\section{Introduction}

Grammar compression aims at reducing the size of an input string $S\in\Sigma^n$ by replacing it with a (small) set of grammar productions $G$ generating $S$ (and only $S$) as output.
Despite generating the smallest such grammar has been proved to be NP-hard~\cite{charikar2005smallest}, several approximation techniques have been developed during the last decades which produce very small grammars on inputs of practical interest. 
Among these techniques, Re-Pair~\cite{larsson2000off} (short for Recursive Pairing) is a simple and fast off-line compression scheme that generates the grammar by recursively replacing high-frequency character pairs with new grammar symbols. Despite its simplicity, Re-Pair achieves high-order entropy compression~\cite{NR2008} and---especially on repetitive datasets---is an excellent compressor in practice~\cite{Wan99,GN2007,CN2010}. This feature makes it the favorite choice in  applications where grammar compression is convenient over other strategies (e.g. compression and indexing of repetitive collections~\cite{GN2007,claude2010compressed,claude2011indexes}). Re-Pair works as follows on a string $S$. As long as there is a pair of adjacent symbols occurring at least twice:

\begin{itemize}
	\item Find the most frequent pair $ab$.
	\item Let $X$ be a new symbol not appearing in $S$. Add the rule $X\rightarrow ab$ and replace all occurrences of $ab$ in $S$ with $X$.
\end{itemize}

Letting $\Sigma=\{0,\dots, \sigma-1\}$, this procedure generates a grammar of size $d$ of the form $X_i \rightarrow Y_iZ_i$, with $X_i\in\{\sigma,\dots, d-1\}$ and $Y_i,Z_i\in \{0,\dots, \sigma+d-1\}$ for $i=0,\dots, d-1$, together with a text $T\in \{0,\dots, \sigma+d-1\}^*$ without repeated character pairs. The first algorithm implementing this strategy was described by Larsson and Moffat in their original Re-Pair paper~\cite{larsson2000off}. This algorithm runs in optimal $\bigO(n)$ time but is very space-consuming, requiring $5n + 4\sigma^2 + 4d + \sqrt n$ words of working space on top of the text. This space does not scale well with the alphabet size, and is particularly high if the input string is not very compressible, i.e. if $d\approx n$. A very fast and more space-efficient implementation of this algorithm exists~\cite{nav2010repair}, and it requires about $12n$ Bytes of main memory during execution. Very recently, this space was reduced---without increasing running times---to $(1+\epsilon)n+\sqrt n$ words (on top of the text stored in $n$ words) for any constant $0<\epsilon \leq 1$ \cite{bille2017space}. This space saving, however, comes at the expenses of practicality: the algorithm in \cite{bille2017space} makes use of complex sub-procedures---such as in-place radix sorting---and is therefore not suitable for a practical direct implementation. The first contribution of this paper is a practical and even more space-efficient variant of the strategy proposed in \cite{bille2017space}. Our algorithm makes use of new techniques of independent interest (such as a very practical integer clustering procedure), runs in linear time, and uses---including the space for storing the text---only $(1.5+\epsilon)n$ words of space during execution for some small constant $\epsilon>0$ (therefore further improving upon \cite{bille2017space}). Our implementation~\cite{prezza2016repair} reduces by half the working space of the state of the art~\cite{nav2010repair}. 

A second concern that should be considered by a good grammar compressor is how to represent the output grammar using the information-theoretic minimum number of bits. 
This problem has recently been addressed in~\cite{tabei2013succinct}, where the authors show that the information-theoretic minimum number of bits needed to represent a grammar with $d$ rules of the form $X\rightarrow ZY$ (the result therefore applies also to Re-Pair) is $\log(d!)+2d\approx d\log d+0.557 d$. In the same paper, the authors show an encoding---based on monotone subsequences decomposition---achieving $d\log d + 2d$ bits in the worst case. Their encoding can be computed in  $\bigO(d^{1.5})$ time. In this paper, we show how to exploit a degree of freedom in the Re-Pair procedure specification in order to maximize the compressibility of the output grammar. Our improved Re-Pair algorithm runs in optimal $\bigO(n)$ time, and the output grammar can be encoded in optimal $\bigO(d)$ time with our technique. We bound the size of our encoding in terms of the number $M\leq d$ of distinct frequencies in the right-productions of the output grammar (i.e. frequencies of pairs at substitution time): our final grammar representation takes at most $d(\log d + \log M + 1) + M\log(d/M) + o(d\log d)$ bits. As we show experimentally, $M$ is orders of magnitude smaller than $d$ on real datasets, making this strategy very effective in practice: on average, our encoding uses just $2.8\%$ more bits than the information-theoretic minimum (with a very small variance and achieving compression in half of the cases). Our  strategy turns out to be very effective when compared with the most efficient compressors used in practice: in half of the tested cases, our compressor improves the output size of \texttt{7-Zip} with maximum compression rate turned on.


\section{Space-Efficient Re-Pair}

In this section we give an overview of the algorithm~\cite{bille2017space} that obtained  $(1+\epsilon)n + \sqrt n$ words of space on top of the text and expected $O(n/\epsilon)$ time, for any constant $0<\epsilon \leq 1$. 

One of the most space-consuming components of Larsson's and Moffat's solution~\cite{larsson2000off} is the text $S$, which is represented as a doubly-linked list of characters to support fast pair replacement. The first idea to save space upon this text representation is to represent the text as a plain word-vector and write blank characters ´$\_$´ in text positions where we replace pairs: when performing replacement $X\rightarrow ab$ at text position $i$, the occurrence of $ab$ starting in $S[i]$ is replaced by $X\_$. Note that, after several replacements, there could be long (super-constant) runs of blanks in the text. To keep operations efficient, in~\cite{bille2017space} we show how to skip such runs in constant time by storing pointers at the beginning and end of each run. This trick gives us constant-time pair extraction from the text.

The idea, at this point, is to insert pairs in a queue and extract them by decreasing frequency. Each time we extract the maximum-frequency pair $AB$, we replace all text occurrences of $AB$ with a new grammar symbol $X$. In order to keep space usage under control, we use two main strategies: (i) we define a frequency cut-off equal to $c\cdot \sqrt n$, for some constant $c$ (the exact value of $c$ is used to keep space under control and is not relevant for this discussion, see ~\cite{bille2017space} for full details), call \emph{high-frequency}  (resp. \emph{low-frequency}) \emph{pairs} those character pairs appearing more (resp. less) than $c\cdot \sqrt n$ times in the text, and use two different queues for high- and low- frequency pairs. (ii)
We keep a position table (array) $\mathrm{TP}$ that, for each pair in the queue, holds the positions of the occurrences of that pair in the text. This position array improves upon the space of the original solution\cite{larsson2000off}, where pairs' occurrences are stored using linked lists.  Let $F_{ab}$ denote the frequency of pair $ab$. The positions of all occurrences of pair $ab$ are stored in a $\mathrm{TP}$-interval $\mathrm{TP}[P_{ab}\ldots P_{ab}+L_{ab}-1]$, and the queue element associated with $ab$ stores values $P_{ab},L_{ab}$, and $F_{ab}$. Invariant $F_{ab} \leq L_{ab} \leq 2F_{ab}$ is valid at all times, and is at the core of an amortization policy guaranteeing efficient operations on the queues in small space (read below).

The space-efficient Re-Pair algorithm works as follows on a string $S$. We start by using the high-frequency queue---we call this \emph{high-frequency phase}---, and then switch to the low-frequency queue when all text pairs have frequency smaller than $c\cdot \sqrt n$---we call this \emph{low-frequency phase}.
\subparagraph*{Algorithm} 
As long as the queue $Q$ is non-empty:
\begin{itemize}
	\item Extract the most frequent pair $AB$ from $Q$.
	\item Let $X$ be a new symbol not appearing in $S$. Output rule $X\rightarrow AB$.
	\item Replace all occurrences of $AB$ in $S$ with $X$ using the position array $\mathrm{TP}$.
	When replacing $xABy$ with $xXy$, decrease the frequencies of $xA$ and $By$ in the queue. 
	
	
	\item For all decreased pairs $cd$ such that $F_{cd} < L_{cd}/2$: update $\mathrm{TP}[P_{cd}\ldots P_{cd}+L_{cd}-1]$ using the amortization procedure \texttt{Synchronize} described below
	\item Update $\mathrm{TP}[P_{AB}\ldots P_{AB}+L_{AB}-1]$ using procedure \texttt{Synchronize} and remove $AB$ from the queue.
\end{itemize}

	\noindent	 \texttt{Synchronize}($ab$): When entering in this procedure, $\mathrm{TP}[P_{ab}\ldots P_{ab}+L_{ab}-1]$ (possibly) contains pairs $xy$ different from $ab$ and/or blanks. The aim is to re-organize the sub-array in such a way that $\mathrm{TP}[P_{ab}\ldots P_{ab}+L_{ab}-1]$ contains only occurrences of $ab$.
	 
	\begin{itemize}
		
		\item Sort $\mathrm{TP}[P_{ab}\ldots P_{ab}+L_{ab}-1]$ by character pairs (ignoring blanks). This procedure uses in-place radix sorting.
		
		\item Compute $F_{xy}$ and $L_{xy}$ for all the pairs $xy$ contained in $\mathrm{TP}[P_{ab}\ldots P_{ab}+L_{ab}-1]$ including $ab$, and insert them in the queue $Q$ (only if their frequency is at least $c\cdot \sqrt n$ if $Q$ is the high-frequency queue)

	\end{itemize}
	
In~\cite{bille2017space} we show that the above amortization policy (i) preserves correctness, in the sense that we always correctly extract the maximum-frequency pair from the queue, and (ii) permits to implement queues operations efficiently.

\subparagraph{Queue operations} 
Both queues need to support the following operations (in addition to \texttt{Synchronize}): 

\begin{itemize}
	\item $\mathcal Q[ab]$: return the triple $\langle P_{ab}, L_{ab}, F_{ab} \rangle$ associated with pair $ab$
	\item $\mathcal Q.remove(ab)$ : remove $ab$ from $Q$
	\item $\mathcal Q.contains(ab)$: return \texttt{true} iff $Q$ contains pair $ab$
	\item $\mathcal Q.decrease(ab)$: decrease $F_{ab}$ by one
	\item $\mathcal Q.insert(ab, P_{ab}, L_{ab}, F_{ab})$: insert $ab$ and its associated information in $Q$
	\item $\mathcal Q.max()$: return the pair $AB$ with largest $F_{AB}$ 
\end{itemize}

The high-frequency queue contains pairs occurring at least $c\cdot \sqrt n$ times, therefore its maximum capacity is $\bigO(\sqrt n)$. 
The high-frequency phase ends when the queue is empty, i.e., when there are no more pairs occurring at least $c\cdot \sqrt n$ times.
In~\cite{bille2017space} we show how to implement all operations on this queue in constant (expected, amortized) time, except \texttt{max} and \texttt{remove}---which are supported in $\bigO(\sqrt n)$ time (i.e. with a simple linear scan of the queue)---and $\texttt{Synchronize}(ab)$---which is supported in $\bigO(L_{ab} + N\cdot \sqrt n)$ time, where $L_{ab}$ is $ab$'s interval length at the moment of entering in this procedure, and $N$ is the number of new pairs $xy$ inserted in the queue. Since we execute at most $\bigO(\sqrt{n})$ times \texttt{max} and \texttt{remove} (once per high-frequency pair), we spend overall $\bigO(n)$ time on the high-frequency queue.


The capacity of the low-frequency queue is $\epsilon\cdot n$, for an arbitrary $0<\epsilon\leq 1$. All operations on this queue run in constant (expected, amortized) time, except  $\texttt{Synchronize}(ab)$---which is supported in $\bigO(L_{ab})$ expected time, $L_{ab}$ being $ab$'s interval length at the moment of entering in this procedure.
We point out and fix a mistake we had in~\cite{bille2017space}. Here, we mistakenly claimed we could implement \texttt{remove} in constant time on the low-frequency queue. Unfortunately, this is not true. We fix this mistake in the next section using amortization (this works also for the original solution): instead of deleting the least occurring pair, we remove the least frequent half of the pairs, when the queue is filled up to its max capacity. This gives amortized constant time for \texttt{remove}.

We fill the low-frequency queue at most $\bigO(1/\epsilon)$ times and hence it follows that our overall algorithm runs in $\bigO(n/\epsilon)$ expected time.

\section{Implementation}

The main differences between our theoretical proposal and the implementation here described are:

\begin{enumerate}
	\item The skippable text representation: our implementation of this component uses only $50\%$ of the space of the theoretical version.
	\item The way we cluster pairs in the $TP$ array: we replace in-place radix sorting with a very efficient and practical in-place clustering algorithm.
	\item The queues implementation: we replace linked lists with more cache-efficient plain vectors of pairs. Moreover, we use a frequency cut-off of $n^{2/3}$ to distinguish between high-frequency and low-frequency pairs (in~\cite{bille2017space} the cut-off was $\bigO(\sqrt n)$). This cut-off allows us to achieve linear running time for our clustering procedures while using sublinear space for the universal tables.  
	\item We show how to produce a more regular Re-Pair grammar and how to efficiently compress it. On average, our encoding uses almost exactly the information-theoretic minimum number of bits.
\end{enumerate}

Since we deal only with the ASCII alphabet, in our implementation we assume that the alphabet size $\sigma$ is constant and, in particular, that we can fit alphabet characters in half a memory word.

\subsection{A more space-efficient skippable text}

We represent the input string $S$ as an array of $n$ half-word locations. Note that grammar symbols can be as large as $n$, so they do not necessarily fit in half a memory word (since we assume $w=\log_2 n$). However we observe that---whenever we perform a pair replacement---the new grammar symbol is always followed by a blank character (because, after a replacement $X\rightarrow AB$, we replace $AB$ with $X\_$): we therefore have available a full word for storing the grammar symbol. We use a word-packed bitvector $M[1,\dots,n]$ to mark non-blank positions with a bit set. If $M[i,i+1]=11$, then the $i$-th text symbol is stored in $S[i]$ (in half a word). If $M[i,i+1]=10$, then the $i$-th text symbol is stored in $S[i,i+1]$ (in a full word). Finally, we skip in constant time runs of blank characters as follows. If the run's length is shorter than or equal to $w$, then we skip the run querying in parallel $w$ bits of $M$ (i.e. extracting the leftmost bit set of some packed size-$w$ sub-array of $M$). Otherwise, we explicitly store the run's length in a word: we keep a word array $W[1,\dots, n/w]$ and store the length of the run starting in position $i$ inside $W[\lfloor i/w\rfloor]$. The total size of our skippable text is $0.5 n + o(n)$ words.

\subsection{Pair clustering and queues implementation}

Note that our algorithm does not actually need the pairs to be \emph{sorted} in the TP array; the correctness of our procedures is preserved if we just \emph{cluster} text positions by character pairs. In this section we show how we achieve this task with a very practical linear-time and in-place procedure. 

We start by defining a frequency cut-off $f=n^{2/3}$. Pairs with frequency larger than $f$ are processed in the high-frequency phase. 
Let $d_{HF}$ be the number of distinct high-frequency pairs. Clearly, $d_{HF}\cdot f \leq n$, so in the high-frequency phase we process at most $d_{HF} \leq n/f = n^{1/3}$ pairs. Let $\Sigma_{HF} = \{0,\dots, \sigma + d_{HF}-1\}$ be the alphabet composed by the original alphabet $\Sigma$ plus grammar symbols created during the high-frequency phase. We have that $|\Sigma_{HF}| < \sigma + n^{1/3}$. This, together with the assumption $\sigma\in\bigO(1)$, implies that a table directly addressing all pairs of symbols in the high-frequency phase contains no more than $|\Sigma_{HF}|^2 = (n^{1/3}+\sigma)^2 = \bigO(n^{2/3})$ entries. The idea, at this point, is to use such tables to cluster the pairs contained in \texttt{TP} sub-arrays in-place (modulo the tables) and in linear time during the high-frequency phase. Algorithm \ref{alg:counting-cluster} reports our clustering procedure. When entering in Algorithm \ref{alg:counting-cluster}, we assume that two tables $C_1, C_2 : \Sigma_{HF} \times \Sigma_{HF} \rightarrow [0,\dots, n)$  have been pre-allocated. $C_1$ is filled with 0, while $C_2$ with NULL values. The algorithm uses these tables to cluster pairs in the input vector $A$ and, before exiting, resets the used entries to 0 (in $C_1$) or NULL (in $C_2$). This is the intuition behind our clustering procedure. After counting pairs (first two \texttt{for} loops in Algorithm \ref{alg:counting-cluster}), $C_1[ab]=C_2[ab]$ store the first position containing $ab$ in the final clustered array. Then, inside our main procedure (\texttt{while} loop in Algorithm \ref{alg:counting-cluster}) we use an index $j$ to scan $A$-positions left to right. The following two invariants are maintained: (i) $A[0,\dots, j-1]$ is clustered, and (ii) $A[C_1[ab],\dots, C_2[ab]-1]$ contains only pairs equal to $ab$. At each cycle, either value $A[j]$ is already in the correct position (i.e. it contains a pair $ab$ such that $C_1[ab] \leq j < C_2[ab]$) and we increment $j$, or it contains a pair $ab$ but $j$ is not inside interval $[C_1[ab],C_2[ab])$. In such case, we place $A[j]$ in the correct position (i.e. $C_2[ab]$) with a swap operation and increment $C_2[ab]$ (since now $A[C_1[ab],\dots, C_2[ab]]$ contains only pairs equal to $ab$). 

Note that (a) $j$ cannot be incremented more than $|A|$ times (\texttt{while} condition), and (b) $C_2[ab]$ cannot be incremented more than $f_{ab}$ times, $f_{ab}$ being the number of occurrences of $ab$ in the text positions contained in $A$ (this is implied by our invariant (ii)). This implies that, in Algorithm \ref{alg:counting-cluster}: (a) Line 14 is executed at most $|A|$ times, and (b) Lines 16-17 are executed at most $\sum_{ab\in A}f_{ab} = |A|$ times (for simplicity $ab\in A$ indicates that pair $ab$ occurs in one of the text positions contained in $A$). Clearly, we spend $\bigO(|A|)$ time inside the three \texttt{for} loops in Algorithm \ref{alg:counting-cluster}. It follows that the whole clustering procedure runs in $\bigO(|A|)$ time. 

\begin{algorithm}[th!]
	\caption{$cluster(A)$}
	\label{alg:counting-cluster}
	
	\SetKwInOut{Input}{input}
	\SetKwInOut{Output}{behavior}
	\SetSideCommentLeft
	\LinesNumbered
	
	\Input{Array $A$ of text positions}
	\Output{Cluster $A$'s entries by character pairs}
	
	\BlankLine
	
	\For{$i=0, \dots, |A|-1$}{
	
		$ab \leftarrow T.pair\_starting\_at(A[i])$\;
		$C_1[ab] = C_1[ab] + 1$\;	
		
	}
	
	\BlankLine
	$j \leftarrow 0$\;
	
	\For{$i=0, \dots, |A|-1$}{
		
		$ab \leftarrow T.pair\_starting\_at(A[i])$\;
		
		\If{$C_2[ab] = NULL$}{
			
			$j\leftarrow j+C_1[ab]$\;
						
			$C_1[ab] \leftarrow C_2[ab] \leftarrow (j - C_1[ab])$\;
			
		}
		
	}
	
	\BlankLine
	$j \leftarrow 0$\;
	
	\While{$j<|A|$}{
	
		$ab \leftarrow T.pair\_starting\_at(A[j])$\;

		\eIf{$C_1[ab] \leq j < C_2[ab]$}{
			
			$j\leftarrow j+1$\;
			
		}{
		
			swap($A, j, C_2[ab]$)\;
			$C_2[ab] \leftarrow C_2[ab]+1$\;
			
		}
		
	}
	
	\BlankLine
	\For{$i=0, \dots, |A|-1$}{
		
		$ab \leftarrow T.pair\_starting\_at(A[i])$\;
		
		$C_1[ab] \leftarrow 0$\;
		$C_2[ab] \leftarrow NULL$\;

	}

\end{algorithm}

\subparagraph{High-frequency queue}\label{sec:HF}

Our high-frequency queue $\mathcal Q_{HF}$ is implemented simply as a matrix (i.e. a direct-access table) $H:\Sigma_{HF} \times \Sigma_{HF} \rightarrow \mathbb N^3$ mapping high-frequency pairs to their coordinates in the \texttt{TP} array: $H[ab] = \langle P_{ab}, L_{ab}, F_{ab} \rangle$ (these variables have the same meaning as in our original theoretical proposal). Then, $\mathcal Q_{HF}[ab]$ requires just an access on $H$,  $\mathcal Q_{HF}.remove(ab)$ requires setting $H[ab]$ to $\langle NULL,NULL,NULL \rangle$, $\mathcal Q_{HF}.contains(ab)$ is implemented by checking whether $H[ab]$ does not contain $NULL$ values, $\mathcal Q_{HF}.decrease(ab)$ requires decrementing the third component of $H[ab]$, and inserting $\langle ab, P_{ab}, L_{ab}, F_{ab} \rangle$ in $\mathcal Q_{HF}$ requires just setting $H[ab] = \langle P_{ab}, L_{ab}, F_{ab} \rangle$.
To compute $\mathcal Q_{HF}.max()$, we just scan every entry of $H$ and return the pair $ab$ with the highest $F_{ab}$. This operation takes $\bigO(|\Sigma_{HF}|^2) = \bigO(n^{2/3})$ time. 

\subparagraph{Low-frequency queue}\label{sec:LF}

Let $\bar\Sigma = \{0,\dots, \sigma+d-1\}$ be the final alphabet including original characters from $\Sigma$ and grammar symbols. 
Let moreover $\epsilon>0$ be an arbitrarily small positive constant. We allow at most $\epsilon\cdot n$ pairs to be in the low-frequency queue at the same time.
Our low-frequency queue $\mathcal Q_{LF}$ is implemented as a quadruple $\langle H, F, max_{f}, ext\rangle$, where:

\begin{itemize}
	\item $H:\bar\Sigma \times \bar\Sigma \rightarrow \mathbb N^3$ is a hash table mapping low-frequency pairs to their coordinates in the $TP$ array: $H[ab] = \langle P_{ab}, L_{ab}, F_{ab} \rangle$. In this case, $H$ is implemented with linear probing and only guarantees expected constant-time operations. We allocate $2\epsilon\cdot n$ slots for $H$ (maximum load factor is $0.5$).
	\item $F : [2,\dots,n^{2/3}] \rightarrow (\bar\Sigma\times \bar\Sigma)^* $ is a vector mapping frequencies of pairs in the queue to (a superset) of all pairs with that frequency: throughout the execution of our algorithm, all pairs with frequency $f$ are stored in vector $F[f]$. However, $F[f]$ may contain also pairs that are no more in the queue or that have a frequency smaller than $f$ (this happens because we amortize operations on $F$, read below).
	\item $max_{f}$ is the frequency of the most frequent pair in the text. Note that this value can only decrease during computation. 
	\item $ext$ is an index such that all pairs in $F[max_{f}][1,\dots, ext]$ have already been extracted from the queue (and thus processed).
\end{itemize}

At this point, $\mathcal Q_{LF}[ab]$ requires just an access on $H$, and $\mathcal Q_{LF}.contains(ab)$ is implemented by checking whether $H$ contains pair $ab$. $\mathcal Q_{LF}.decrease(ab)$ is implemented with a ``lazy'' strategy: let $f$ be $ab$'s frequency.  We append $ab$ at the end of $F[f-1]$, we decrement the third component of $H[ab]$, but we do not remove $ab$ from  $F[f-1]$ (this would require time proportional to $|F[f-1]|$). However, we keep $|F|$ counters storing the number of deleted pairs in each $F[i]$, and rebuild $F[i]$ whenever the number of deleted pairs is above $|F[i]|/2$. The rebuilding process requires removing from $F[i]$ pairs that are no longer in $H$ or whose frequency is no longer $i$ (this can be checked accessing $H$). It is easy to see that---thanks to this amortization strategy---rebuilding $F$'s elements adds no asymptotic cost to our procedures.
To insert $\langle ab, P_{ab}, L_{ab}, F_{ab} \rangle$ in $\mathcal Q_{LF}$, we set $H[ab] = \langle P_{ab}, L_{ab}, F_{ab} \rangle$ and append $ab$ at the end of $F[F_{ab}]$. 
To compute $\mathcal Q_{LF}.max()$, we access $F[max_f][ext]$ for increasing values of $ext$ until we find a pair that is in $H$ and has frequency $max_f$. If $ext$ reaches the end of $F[max_f]$, we decrement $max_f$ by one, reset $ext$ to 0, and proceed with the search (note: $ext$ needs to be reset to 0 also after rebuilding $F[max_f]$). Note that, thanks to our amortization policy on $F$ and to the fact that $max_f$ can only decrease, the overall time spent inside $\mathcal Q_{LF}.max()$ cannot exceed $\bigO(n)$.
We can deal with queue overflows (i.e. cases where we insert more than $\epsilon\cdot n$ pairs in  $\mathcal Q_{LF}$) as follows. We keep a counter storing the number of pairs in the queue (updating it each time we either extract the maximum or insert a new pair). Whenever this counter reaches size $\epsilon\cdot n$, we remove from  $\mathcal Q_{LF}$ the $0.5\epsilon\cdot n$ pairs with the lowest frequency. It is easy to see that this operation takes amortized constant time over all insert operations on the queue.


\subparagraph{Analysis}

All operations in the high-frequency queue take constant amortized time except \texttt{max()}, which takes $\bigO(n^{2/3})$ time. Note that we call \texttt{max()} at most $d_{HF}\leq n^{1/3}$ times (i.e. the maximum number of high-frequency pairs), therefore the overall time spent inside this function is $\bigO(n)$. We perform a constant number of operations for each text occurrence of a high-frequency pair. Being the overall number of occurrences of high-frequency pairs $\bigO(n)$, it follows that we spend overall $\bigO(n)$ time on the high-frequency queue.

In the low-frequency queue, all operations take constant amortized expected  time. Note that we may need to re-fill the queue (up to) $\bigO(1/\epsilon)$ times, being its capacity $\bigO(\epsilon\cdot n)$. Again, the total number of occurrences of low-frequency pairs cannot exceed $n$ and we perform a constant number of operations on each pair occurrence every time the queue is re-filled, therefore we spend overall $\bigO(n/\epsilon)$ expected time on the low-frequency queue; this time dominates the overall time of our algorithm.

Our high-frequency queue takes $\bigO(|\Sigma_{HF}|^2) = \bigO(n^{2/3})$ words of space, while our low-frequency queue takes $\epsilon\cdot n$ words of space for an arbitrarily small constant $\epsilon$. Taking into account all components, the overall space used in RAM by our practical implementation is (including the skippable text) $(1.5+\epsilon)n$ words, for some small positive constant $\epsilon$ (where we hide $o(n)$ terms inside $\epsilon n$). 

\subsection{Compressing the final grammar}\label{sec:compress grammar}

The problem of succinctly representing  straight line programs (SLPs) has lately been addressed in~\cite{tabei2013succinct}. In this work, the authors propose an encoding of at most $d\log d + 2d$ bits for a grammar consisting of $d$ symbols. This bound is very close to the information-theoretic lower bound of $\log(d!)+2d \approx d\log d + 0.557d$ bits (also proved in the same paper), and considerably improves the straightforward encoding of $2d\log d$ bits obtained by explicitly storing the two grammar symbols of each production. One of the drawbacks of this solution is, however, its running time of $\bigO(d^{1.5})$, which can become prohibitively high if the input text is not very compressible (i.e. $d\approx n$). In this paragraph, we show how to exploit a degree of freedom in the Re-Pair algorithm specification (namely, the processing order of same-frequency pairs in the queue) in order to produce a more compressible grammar. Our Re-Pair variant runs with the same space/time bounds of our original proposal described in the previous sections. The algorithm for compressing the output grammar runs in optimal $\bigO(d)$ time and, on average, compresses the grammar with only a negligible overhead on top of the optimal $\log(d!)+2d$ bits (see experimental section).

Our strategy is the following. First of all note that, when choosing the pair with the highest frequency $max_f$ from our queues, this pair is not univoquely determined (as there can be more than one pair with frequency $max_f$). Our idea is to choose the extraction order of pairs with frequency $max_f$ in such a way that the output sequence of grammar productions is more compressible. We extract max-frequency pairs in increasing order according to the ordering $\prec_{max}$ defined as $ab \prec_{max} cd \Leftrightarrow max(a,b)<max(c,d)$. In the high-frequency queue, the maximum is extracted scanning all pairs in the queue; it follows that the extraction order can be easily modified to that defined by $\prec_{max}$ without affecting running times. In the low-frequency queue, as soon as we decrease $max_f$, we sort the pairs in $F[max_f]$ according to $\prec_{max}$. In theory, this task can easily be achieved in linear time and in-place by using in-place radix sorting~\cite{FMP2007}. In our implementation, considering that the pairs in $F[max_f]$ are stored contiguously in main memory (i.e. cache locality can be exploited), we use C++ std's  \texttt{sort}. At this point, note that the sequence of pairs extracted (left-to-right) from $F[max_f]$ is ordered according to $\prec_{max}$ \emph{provided} that $F[max_f]$ is not modified (i.e. no pairs are added to it) after the sorting procedure. Note that we either remove or append pairs in $F[max_f]$. Removing pairs does not invalidate the ordering. We append some pair at the end of $F[max_f]$ only when we extract a pair $ab$ from the queue and $ab$ is always followed (resp. preceded) by the same character $c$. In this case, after replacement $X\rightarrow ab$, a new pair $Xc$ (resp, $cX$) with frequency $max_f$ appears in the text and is therefore appended at the end of $F[max_f]$. However, note that $X$ is (by definition) larger than all symbols appearing in the text; it follows that order $\prec_{max}$ is preserved when appending $Xc$ (resp, $cX$) at the end of $F[max_f]$. Let $M$ be the number of \emph{distinct frequencies} in the right-productions of the output grammar (i.e. frequencies of pairs at substitution time). Note that $M$ can be much smaller than $d$ (one to three orders of magnitude on real-case examples, as shown in the experimental section). We encode the output grammar $G = \{\sigma \rightarrow a_1b_1, (\sigma+1) \rightarrow a_2b_2, \dots, (\sigma+d-1) \rightarrow a_db_d\}$  as follows. We delta-encode (using Elias delta-encoding) the (at most) $M$ increasing sub-sequences of $max( a_1, b_1),  \dots, max( a_d, b_d)$ in $d\log M (1+o(1))$ bits; we encode the lengths of these sub-sequences in $M\log(d/M)(1+o(1))$ bits;  we store with Elias delta-encoding the sequence $|a_1 - b_1|, \dots, |a_d - b_d|$ in $d\log d(1+o(1))$ bits (note that this is a very pessimistic upper bound); we store a length-$d$ bitvector recording whether $a_i > b_i$, for $i=1,\dots, d$. In total, our encoding uses at most $d(\log d + \log M + 1) + M\log(d/M) + o(d\log d)$ bits. 


\section{Experimental results}

We compared running times, memory usage, and compression rates of our~\cite{prezza2016repair} (\texttt{rp}) and Navarro's~\cite{nav2010repair} (\texttt{NAV}) implementations of Re-Pair against the compression tools \texttt{7-Zip}~\cite{7zip} (with maximum compression rate turned on, i.e. using option \texttt{-mx=9}), and \texttt{bzip2}~\cite{bzip2}.
We ran all tools on two artificial extremely repetitive datasets---\texttt{fib41} and \texttt{tm29}---, on three real repetitive datasets---\texttt{boost}, \texttt{cere}, and \texttt{einstein}---, and on three real not-so-repetitive datasets---\texttt{dblp}, \texttt{english}, and \texttt{sources}. All datasets except \texttt{boost} come from the \texttt{pizza\&chili} corpus~\cite{pizzachili}. Dataset \texttt{boost} consists of concatenated versions of the C++ \texttt{boost} library~\cite{boost}, a collection that turns out ot be very repetitive.
We moreover compressed with our tool all \texttt{pizza\&chili} real datasets---truncated to 50MB when bigger---and compared the compressed files' sizes with the information-theoretic minimum number of bits needed to represent them.
Recall that Re-Pair outputs a grammar consisting of $d$ productions plus a text $T\in\bar\Sigma^t$, where $\bar\Sigma = \{0,\dots, \sigma+d-1\}$, such that every $XY\in \bar\Sigma^2$ appears at most once in $T$. It follows, from~\cite{tabei2013succinct}, that the minimum number of bits to represent the compressed file is $\log(d!)+2d+t\log(\sigma+d) \approx d(\log d + 0.557) + t\log(\sigma+d)$.

\begin{table}
	\centering
	\begin{tabular}{|l|l|l|l|l|l|}\hline
		&plain	&\texttt{7-Zip}	& \texttt{bzip2}	&\texttt{NAV}	& \texttt{rp}\\\hline
		boost	&800.00	&0.162	&51.87	&0.243	&0.087\\
		cere	&439.92	&5.00	&110.96	&22.04	&8.50\\
		dblp	&200.00	&21.90	&22.71	&42.58	&23.91\\
		einstein	&800.00	&1.16	&45.50	&3.16	&1.46\\
		english	&800.00	&194.29	&226.60	&337.14	&181.55\\
		fib41	&255.50	&0.450272	&0.014203	&0.000307	&0.000044\\
		sources	&200.00	&29.78	&37.32	&77.06	&42.61\\
		tm29	&256.00	&0.91866	&0.031403	&0.000601	&0.000132\\\hline
	\end{tabular}\caption{Size of the uncompressed (column \emph{plain}) and compressed files. Space is expressed in MB. In half of the cases (the most repetitive ones), \texttt{rp} improves the compression rate of \texttt{7-Zip}.}\label{tab:sizes}
\end{table}

\begin{table}
	\begin{tabular}{|l|l|l|l|l||l|l|l|l|}\hline
		&\texttt{7-Zip}	(C) &\texttt{bzip2}	(C)&\texttt{NAV}	(C)&\texttt{rp} (C)&\texttt{7-Zip} (D) &\texttt{bzip2}	(D)&\texttt{NAV}	(D)&\texttt{rp} (D)\\\hline
		boost	&683	&8	&9602	&4450&68	&4	&1	&6\\
		cere	&683	&8	&5281	&2838&70	&4	&20	&136\\
		dblp	&683	&7	&2405	&1136&69	&4	&13	&162\\
		einstein	&683	&8	&9606	&4381&70	&4	&3	&17\\
		english	&683	&7	&9606	&5755&69	&4	&168	&1239\\
		fib41	&683	&8	&3068	&1665	&69	&4	&1	&5\\
		sources	&683	&8	&2407	&1572&69	&4	&32	&289\\
		tm29	&683	&8	&3074	&1667&69	&4	&1	&5\\\hline
	\end{tabular}\caption{RAM usage of the tools executed on the files of Table \ref{tab:sizes} during compression (C) and decompression (D). Space is expressed in MB.}\label{tab:ws}
\end{table}

\begin{table}
	\begin{tabular}{|l|l|l|l|l||l|l|l|l|}\hline
		&\texttt{7-Zip}	(C) &\texttt{bzip2}	(C)&\texttt{NAV}	(C)&\texttt{rp} (C)&\texttt{7-Zip} (D) &\texttt{bzip2}	(D)&\texttt{NAV}	(D)&\texttt{rp} (D)\\\hline
		boost	&212.86	&104.76	&179.95	&1040.54&1.20	&12.63	&12.28	&8.07\\
		cere	&525.77	&44.00	&82.56	&3817.68&0.94	&14.15	&7.83	&8.28\\
		dblp	&135.40	&26.35	&73.24	&1134.84&1.43	&4.40	&3.01	&3.61\\
		einstein	&154.15	&124.92	&196.49	&2118.36&1.53	&16.06	&13.60	&9.22\\
		english	&833.08	&87.81	&458.19	&25850.62&10.36	&29.21	&15.56	&25.32\\
		fib41	&36.50	&97.94	&25.36	&71.97&0.34	&4.34	&2.50	&1.47\\
		sources	&117.99	&21.63	&81.46	&3459.01&1.88	&4.66	&3.54	&5.44\\
		tm29	&47.88	&166.41	&24.23	&101.32&0.56	&4.43	&2.54	&1.58\\\hline
	\end{tabular}\caption{Compression (C) and decompression (D) times of the tools executed on the files of Table \ref{tab:sizes}. Times are expressed in seconds.}\label{tab:times}
\end{table}

\begin{table}
	\centering
	\begin{tabular}{|l|l|l|l|l|l|l|}\hline
dataset	&d	&M	&plain	&lower bound	&\texttt{\texttt{rp}}	&rate (\%)\\\hline 
cere	&1712283	&1441	&50	&5.02	&4.60	&91.69	\\
coreutils	&1265102	&2684	&50	&3.76	&3.78	&100.53	\\
dblp	&517857	&1995	&50	&6.16	&6.24	&101.30	\\
dna	&571507	&1551	&50	&15.36	&13.89	&90.42	\\
einstein.de	&35042	&2720	&50	&0.09	&0.11	&128.73	\\
einstein.en	&36707	&2695	&50	&0.09	&0.12	&128.67	\\
english	&2137984	&2375	&50	&12.90	&11.68	&90.58	\\
influenza	&299721	&2288	&50	&1.58	&1.65	&104.00	\\
kernel	&778089	&2617	&50	&2.11	&2.18	&103.39	\\
proteins	&1853756	&2106	&50	&26.06	&24.33	&93.36	\\
sources	&1232862	&2584	&50	&11.64	&11.37	&97.66	\\
pitches	&2480114	&2514	&50	&24.18	&22.73	&94.00	\\
world\_leaders	&209283	&1670	&45	&0.68	&0.77	&112.39	\\\hline
	\end{tabular}\caption{Number of grammar rules $d$, number $M$ of distinct pair frequencies at substitution time, uncompressed file size (column \emph{plain}), information-theoretic lower bound for storing the grammar and the final compressed text, size of our compressed file (column \texttt{\texttt{rp}}), and compression rate of our succinct representation with respect to the information-theoretic lower bound. Space is expressed in MB. On average, our representation's size is only $2.8\%$ larger than the information-theoretic lower bound.}\label{tab:low_bound}
\end{table}

Table \ref{tab:sizes} reports the compressed file's sizes. \texttt{rp} compresses better than \texttt{NAV}: this is expected, since \texttt{NAV} does not compress the output grammar (i.e. 32-bits integers are used to represent grammar rules and the final text). Our tool improves in almost all cases the output of \texttt{bzip2}; this is also expected, considering the small maximum window size of this tool (900 KB). \texttt{7-Zip} compresses better than \texttt{rp} in half of the cases. This is mainly due to the large dictionary size of \texttt{7-Zip} and to the fact that LZ77 is inherently more powerful than grammar compression. On very repetitive files (\texttt{boost}, \texttt{fib41}, and \texttt{tm29}) and on dataset \texttt{english}, however, \texttt{rp}'s compression rate is much better than that of \texttt{7-Zip}. 

Working space of all tools during compression and decompression are reported in Table \ref{tab:ws}. Columns \texttt{NAV} (C) and \texttt{rp} (C) show that our goal of improving the state of the art's working space is achieved: our tool reduces by a factor of 2 \texttt{NAV}'s memory requirements. Note that the tools \texttt{7-Zip}	and \texttt{bzip2} compress the file in windows of fixed size and therefore use a constant (and much smaller) working space. As far as decompression is concerned,  \texttt{rp} uses a very variable working space across the datasets, with very repetitive files being decompressed in much less space than less repetitive ones. In all cases, \texttt{rp} uses much less working space during decompression than compression; this space is about one order of magnitude larger than that of \texttt{NAV}. 

The efficiency of our compressor in terms of compression rate and working space is paid in terms of running times. As shown in Table \ref{tab:times} (first 5 columns), \texttt{rp} is one to two orders of magnitude slower than \texttt{NAV}, despite their theoretical running times being the same (linear). We suspect this is due to the large number of calls to our clustering procedure on the array \texttt{TP} and to our amortization policies (which require to re-build our structures from time to time in order to keep space usage under control). Needless to say, the highly-optimized compressors \texttt{7-Zip}	and \texttt{bzip2} outperform \texttt{rp} in terms of running times (except on extremely high repetitive datasets). On the other hand, our decompressor turns out to be very fast: \texttt{rp} decompresses faster than \texttt{bzip2} and in comparable times with respect to \texttt{NAV}  and---except on very repetitive datasets---\texttt{7-Zip}. On very repetitive datasets, \texttt{rp}'s decompressor is one order of magnitude slower than \texttt{7-Zip}.

To conclude, Table \ref{tab:low_bound} displays the compression rate of our tool on 13 real datasets from the \texttt{pizza\&chili} corpus. The table shows that the number $M$ of distinct pair frequencies at substitution time is one to three orders of magnitude smaller than the number $d$ of rules; this justifies the grammar compression strategy introduced in Section \ref{sec:compress grammar}. In column 5 we show the information-theoretic minimum number of MB needed to store the compressed file, which can be directly compared to the size of our output (column 6). In the last column we report the efficiency of our compressed representation with respect to the information-theoretic minimum. It is very interesting to observe that our representation gets always very close to the lower bound, improving it (i.e. achieving compression) in half of the cases. On average, we use just $2.8\%$ more space than the lower bound. Interestingly, our grammar representation achieves better compression on less compressible files, rather than on those that result in a very small grammar.

\bibliographystyle{abbrv}
\bibliography{paper}

\end{document}